\documentclass[conference]{IEEEtran}

\usepackage{graphicx}

\begin{document}

\title{\normalsize PRECISION STUDY OF HYPERFINE STRUCTURE IN SIMPLE
ATOMS}

\author{\authorblockN{S. Karshenboim\authorrefmark{1}\authorrefmark{2},
S. Eidelman\authorrefmark{3}, P. Fendel\authorrefmark{1}, V. G.
Ivanov\authorrefmark{4}\authorrefmark{2}, N.
Kolachevsky\authorrefmark{5}\authorrefmark{1}, V.
Shelyuto\authorrefmark{2}, and T.~W.~H\"{a}nsch\authorrefmark{1} }
\authorblockA{\authorrefmark{1}Max-Planck-Institut f\"ur Quantenoptik,
85748 Garching, Germany}
\authorblockA{\authorrefmark{2}D. I. Mendeleev Institute for Metrology (VNIIM),
 St. Petersburg 190005, Russia}
\authorblockA{\authorrefmark{3}Budker Institute for Nuclear Physics
and Novosibirsk University, Novosibirsk, Russia}
\authorblockA{\authorrefmark{4}Pulkovo Observatory, St. Petersburg 196140,
Russia}
\authorblockA{\authorrefmark{5}P.N. Lebedev Physics Institute, Moscow, 119991,
Russia} }

\maketitle

\begin{abstract}
We consider the most accurate tests of bound state QED theory of
the hyperfine splitting in two-body atoms related to the HFS
interval of the $1s$ state in muonium and positronium and the $2s$
state in hydrogen, deuterium and the helium-3 ion. We summarize
their QED theory and pay special attention to involved effects of
strong interactions and to recent optical measurements of the $2s$
HFS interval in hydrogen and deuterium. We present results for
specific ratios of the $1s-2s$ frequencies in hydrogen and
deuterium which happen to be among the most accurately measured
and calculated quantities.
\end{abstract}

\section{Introduction}

Recent progress in precision optical measurements provides as with
new accurate data related to simple atoms. The data are related to
such quantities as the Lamb shift and hyperfine intervals.
Previously such data were available only from microwave
measurements. Precision atomic theory is bound state quantum
electrodynamics.

Quantum electrodynamics (QED) by inself is a well established
theory. It successfully describe pure leptonic atoms and partly
conventional atoms such as hydrogen. Problems with the accuracy of
theoretical predictions arise because QED is incomplete in a
sense. Even pure leptonic systems are not free of hadronic effects
which enter through virtual hadronic intermediate states. In case
of hydrogen, deuterium etc we should take into account
distribution of the electric charge, magnetic moment and some more
complicated effects. Here we consider QED tests involving the
hyperfine splitting in light hydrogen-like atoms.

Bound state QED is much more complicated than QED for free
particles and deserves serious tests. Some of such tests are
significant for the determination of fundamental constants and in
particular for the fine structure constant $\alpha$, which may be
obtained from the hyperfine structure (HFS) interval in muonium
(see reviews \cite{PRep1,PRep2,codata} for detail).

The HFS interval in hydrogen and some other light atoms has been
known for a while with a record experimental accuracy, however,
the related theory suffers from uncertainties of the nuclear
structure effects at the one-ppm level or higher.

We consider a few possibilities to perform QED tests going far
beyond this level of accuracy. Some of them are of metrological
interest.

The quantities of interest for the QED tests, such as the Lamb
shift and the HFS intervals, lie in the microwave domain and to
perform an optical measurement for such a quantity one has to deal
with a number of optical quantities and to combine their
frequencies to extract a microwave value of interest. Some
combinations of this kind can be determined with an extremely high
accuracy because in differential measurements some systematic
effects can be cancelled.

In particular, we present here results for specific ratios of the
$1s-2s$ transition frequencies in hydrogen and deuterium
\begin{eqnarray}
R_{\rm HFS}({\rm 1s-2s, H}) &=& \frac{f({\rm 1s-2s},
F=0)}{f({\rm 1s-2s}, F=1)}\;,\\
R_{\rm HFS}({\rm 1s-2s, D}) &=& \frac{f({\rm 1s-2s},
F=1/2)}{f({\rm 1s-2s}, F=3/2)}\;,
\end{eqnarray}
which happen to be among the most accurately measured and
calculated quantities

\section{Experimental determination of the ratio $R_{\rm HFS}({\rm 1s-2s})$ in hydrogen and deuterium}

Some time ago a long-term program was launched at
Max-Planck-Institut f\"ur Quantenoptik to built a new natural
frequency standard locked to the ultraviolet $1s-2s$ transition
\cite{1s2s}. Applying the hydrogen spectrometer developed for this
program we performed differential measurements  for hyperfine
different components of the $1s-2s$ transition frequency in
hydrogen \cite{H2s} and deuterium \cite{D2s} (see Fig.~1 for the
level scheme in the deuterium atom).

\begin{figure}
\begin{center}
\includegraphics[width=6cm]{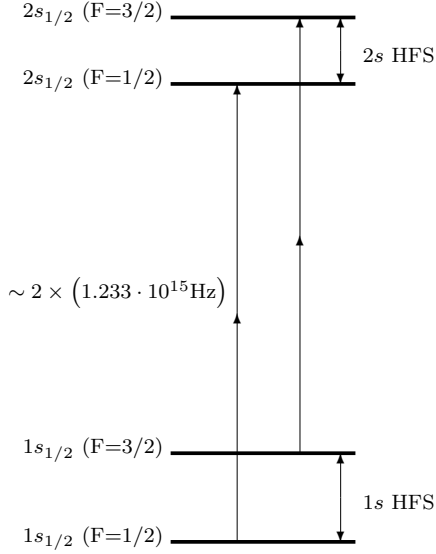}
\bigskip
\caption{Energy levels scheme in deuterium\label{fig}}
\end{center}
\end{figure}

The ratios under question can be present in the form
\begin{eqnarray}
R_{\rm HFS}({\rm 1s - 2s,\;H}) &=& 1 +\frac{ \Delta_{\rm HFS}({\rm
H})
}{f({\rm 1s - 2s}, F=1)}\;,\\
R_{\rm HFS}({\rm 1s - 2s,\;D}) &=& 1 + \frac{\Delta_{\rm HFS}({\rm
D})}{f({\rm 1s - 2s}, F=3/2)}\;,
\end{eqnarray}
where
\begin{eqnarray}
\Delta_{\rm HFS}({\rm H})&=&{f({\rm 1s - 2s},
F=0)-f({\rm 1s - 2s}, F=1)}\;,\nonumber\\
\Delta_{\rm HFS}({\rm D})&=&{f({\rm 1s - 2s}, F=1/2)-f({\rm 1s -
2s}, F=3/2)} \nonumber
\end{eqnarray}
are the subjects of measurements.

The results of our measurements \cite{H2s,D2s,JETP}
\begin{eqnarray}
\Delta_{\rm HFS}({\rm H})&=&1242\,848\,892(16) \;{\rm
Hz}\;,\label{e:dealtaH}\\
\Delta_{\rm HFS}({\rm D})&=&286\,459\,899(7) \;{\rm
Hz}\label{e:dealtaD}
\end{eqnarray}
relate to the following values for the ratio
\begin{eqnarray}
R_{\rm HFS}^{\rm exp}({\rm 1s-2s, H}) &=& 1 +
5039\,813\,857(65)\times10^{-16}\;,\nonumber\\
R_{\rm HFS}^{\rm exp}({\rm 1s-2s, D}) &=& 1 +
1161\,293\,109(28)\times10^{-16}\;,\nonumber
\end{eqnarray}
with the uncertainty completely determined by the experimental
accuracy in the determination of $\Delta_{\rm HFS}$.

The results for the $2s$ hyperfine interval in hydrogen and
deuterium obtained from (\ref{e:dealtaH}) and (\ref{e:dealtaD})
are found to be \cite{H2s,D2s}
\begin{eqnarray}
f_{\rm HFS}^{\rm H}(2s) &=& 177\,556\,860(16)\;{\rm
Hz}\;,\\
f_{\rm HFS}^{\rm D}(2s) &=& 40\,924\,454(7)\;{\rm Hz}\;.
\end{eqnarray}

The hyperfine interval of the $2s$ state was measured by microwave
spectroscopy in hydrogen \cite{H2s_old} and deuterium
\cite{D2s_old} a long time ago. Only recently (in 2000) the
hydrogen result was somewhat improved \cite{H2s_new}. Our optical
results \cite{H2s,D2s} agree well with the early microwave data
mentioned above and are about three times as precise.

\section{Theoretical determination of the ratio $R_{\rm HFS}({\rm 1s-2s})$ in hydrogen and deuterium}

For a theoretical determination of the ratio $R_{\rm HFS}({\rm
1s-2s})$ it is helpful to re-write the ratios in the form
\begin{eqnarray}
R_{\rm HFS}({\rm 1s-2s,\; H}) &=& 1 + \frac{7}{8}\frac{f_{\rm HFS}
({\rm 1s,\; H})}{f({\rm 1s-2s}, F=1)} \nonumber\\&-&
\frac{1}{8}\frac{D_{\rm 21}
({\rm H})}{f({\rm 1s-2s}, F=1)}\;,\\
R_{\rm HFS}({\rm 1s-2s,\; D}) &=& 1 + \frac{7}{8}\frac{f_{\rm HFS}
({\rm 1s,\; D})}{f({\rm 1s-2s}, F=3/2)} \nonumber\\&-&
\frac{1}{8}\frac{D_{\rm 21} ({\rm D})}{f({\rm 1s-2s}, F=3/2)}\;.
\end{eqnarray}
While the frequency of the hyperfine interval of the $1s$ state,
$f_{\rm HFS} ({\rm 1s})$, is well-known experimentally
\begin{eqnarray}
f_{\rm HFS} ({\rm 1s,\;H}) &=& 1420\,405\,751.768(1)\; {\rm Hz}\;,
\nonumber\\
 f_{\rm HFS} ({\rm 1s,\; D}) &=& 327\,384\,352.522(2)\; {\rm
 Hz}\;,
\nonumber
\end{eqnarray}
the specific difference
\begin{equation}
D_{21} = {2^3}\cdot f_{\rm HFS}(2s)-f_{\rm HFS}(1s)
\end{equation}
is a subject of theoretical investigations.

As was shown recently \cite{d21first}, an accurate QED test
competitive with other QED tests on the hyperfine theory  is
possible with ordinary hydrogen atoms. To reach a sensitivity to
higher-order QED corrections, one has to combine the HFS intervals
of the $1s$ and $2s$ states in the same atom in the form of
$D_{21}$. That eliminates the leading nuclear contributions.

A substantial cancellation of the nuclear structure contributions
takes place because the nuclear contribution in leading
approximation is of a specific factorized form of
\begin{equation}
\Delta E({\rm Nucl}) = {A({\rm Nucl})} \times {\big\vert\Psi_{nl}
({\bf r}=0)\big\vert^2}\;.
\end{equation}
In other words, the correction is a product of a certain
nuclear-structure parameter {$A({\rm Nucl})$} and the squared
value of the wave function at the origin
\begin{equation}
{ \big\vert\Psi_{nl}({\bf r}=0)\big\vert^2 = \frac{1}{\pi}
\left(\frac{Z\alpha\,m_R c}{ {n} \hbar} \right)^{{3}} }\delta_{l0}
\;,
\end{equation}
with $n,l$ being the principal and orbital quantum numbers,
respectively, and $m_R$ being the reduced mass of the electron.

Still there are higher-order corrections due to the nuclear
effects which are of a more complicated form and survive this
cancellation. However, they are much smaller and under control
\cite{th2s}. The theory of $D_{21}$ in light hydrogen-like atoms
is presented in Table~\ref{T:th} \cite{th2s,th2s1}.

Here we updated results for the one-loop correction by applying
the extrapolation procedure developed in \cite{th2s1,JETP} to
recently obtained numerical data related to hydrogen-like atoms
for $Z\geq 5$. Our one-loop effective coefficients are found to be
$C_{\rm SE}(Z=1)=2.4(4)$ and $C_{\rm SE}(Z=2)=2.3(2)$ in notation
of \cite{th2s1}.

\begin{table}[hbtp]
\begin{center}
\begin{tabular}{lccc}
\hline
Contribution to HFS in & Hydrogen & Deuterium & $^3$He$^+$ ion\\
& [kHz] & [kHz] & [kHz] \\
\hline
$D_{21}({\rm QED3})$ & 48.937 &  11.305\,6 & -1\,189.253\\
$D_{21}({\rm QED4})$ & {0.022(3)} & {0.004\,7(6)}  &-1.19(5)\\
$D_{21}({\rm Nucl})$ & {-0.002} & {0.002\,6(2)} & 0.307(35)\\
\hline
$D_{21}({\rm theo})$ & 48.955(3) &  11.312\,8(6) & -1\,190.14(6)  \\
\hline
\end{tabular}
\end{center}
\caption{Theory of the specific difference $D_{21}=8f_{\rm
HFS}(2s)-f_{\rm HFS}(1s)$ in light hydrogen-like atoms. The
numerical results are presented for the related frequency
$D_{21}/h$. See \cite{JETP} for notation.\label{T:th}}
\end{table}

With the help of the theoretical results for the specific
difference $D_{21}$ in hydrogen and deuterium, described above, we
arrive at theoretical predictions for the ratio $R_{\rm HFS}({\rm
1s-2s})$ in hydrogen and deuterium
\begin{eqnarray}
R_{\rm HFS}^{\rm th}({\rm 1s-2s, H}) &=& 1 +
5039\,813\,945.3(1.4)\times10^{-16}\;,\nonumber\\
R_{\rm HFS}^{\rm th}({\rm 1s-2s, D}) &=& 1 +
1161\,293\,091.9(0.3)\times10^{-16}\;.\nonumber
\end{eqnarray}
The uncertainty of the theoretical evaluation of $D_{21}$
dominates in the uncertainty budget of $R_{\rm HFS}^{\rm th}$.

\section{Summary of theoretical and experimental determination of the ratio $R_{\rm HFS}({\rm 1s-2s})$
in hydrogen and deuterium}

Recently there have been a number of discussions on most precisely
known quantities from an experimental or a theoretical point of
view. It is indeed quite tricky when one discusses a fractional
accuracy because in experiment and theory the accurate value can
be often achieved when the dominant contribution into a certain
quantity is known and a subject of determination is a difference
between a real value and a dominant contribution. An example is,
e.g, so-called measurements of the Lamb shift in the ground state
in the hydrogen atom, where two transitions, with frequencies
different by a factor approximately equal to four, were compared
\begin{eqnarray}
\frac{4f({\rm 2s-4s})}{f({\rm 1s-2s})} &=& 1 +
7\,781\,376(16)\times10^{-12}\;,~~~\cite{1sL1}, \nonumber\\
\frac{4f({\rm 2s-6d}_{5/2})}{f({\rm 1s-3s})}&=&
1 +6\,431\,080(14)\times10^{-12}\;,~~~\cite{1sL2},\nonumber\\
\frac{4f({\rm 2s-4p}_{3/2})}{f({\rm 1s-2s})}&=&1
+9\,789\,493(16)\times10^{-12}\;,~~~\cite{1sL3}.\nonumber
\end{eqnarray}

We consider here a specific ratio of the $1s-2s$ transition
frequencies in hydrogen and deuterium which offers one of the most
accurate values in both theoretical and experimental lists. The
highest accuracy was also achieved by a splitting of the ratio
into well-established large contributions and small corrections
left to determine experimentally or theoretically.

The results on the determination of $R_{\rm HFS}({\rm 1s-2s})$ in
hydrogen and deuterium are summarized in Table~\ref{tab:R}, which
shows that measurements and calculations of the HFS ratio of the
$1s-2s$ transition frequencies have record fractional accuracy and
the theory is in a good agreement with the experiment.
Nevertheless, one should not overestimate the importance of such
an accuracy. From experimental point of view the most ambitious
task is to obtain the highest accuracy for a directly measured
value, it is similar for theory. What is more important is which
effects could affect the accuracy. In our case, for the experiment
those are spin-dependent effects in the absolute $1s-2s$
measurement. Theoretical effects which limit the accuracy of our
calculations related to our understanding of so-called
state-dependent part of the fourth-order recoil corrections and of
the one-loop and two-loop self-energy contributions. Both
theoretical and experimental problems mentioned are actual
problems to be studied.

\begin{table}
\begin{center}
\begin{tabular}{ccrc}
\hline Atom & Method & Value~~~~~~~~~~~~~~~~ &  $u_r$\\
\hline H & exp & $1 +
5039\,813\,857(65)\times10^{-16}$&$65\times10^{-16}$\\
H & theory & $1 +
5039\,813\,945.3(1.4)\times10^{-16}$&$1.4\times10^{-16}$\\
D & exp & $1 +
1161\,293\,109(28)\times10^{-16}$&$28\times10^{-16}$\\
D & theory & $1 +
1161\,293\,091.9(0.3)\times10^{-16}$&$0.3\times10^{-16}$\\
\hline
\end{tabular}
\caption{Theoretical and experimental values of ratio $R_{\rm
HFS}({\rm 1s-2s})$ in hydrogen and deuterium and their fractional
uncertainty, $u_r$} \label{tab:R}
\end{center}
\end{table}

\section{Other tests of bound state QED related to the hyperfine effects}

\subsection{The difference $D_{21}$ in atoms other than hydrogen and deuterium}

The $1s$ and $2s$ hyperfine intervals are measured much more
accurately compared to a theoretical prediction which can be made
for each of them separately. However, for the difference $D_{21}$
the experimental and theoretical accuracy become competitive.
While for $^3$He$^+$ the experiment \cite{He2s} is already really
competitive with the theory \cite{th2s1}, in the case of hydrogen
and deuterium, the theoretical uncertainty is up-to-date
substantially better than the uncertainty related to the
measurements of the HFS interval in the $2s$ state. Among other
QED tests with $D_{21}$ the case of the helium-3 ion is most
sensitive to various higher-order bound state QED effects. That is
partly because of a higher fractional accuracy in the
determination of the $2s$ hyperfine interval
\cite{He2s}\footnote{The minus sign in the case of helium-3 ion
reflects the fact that the direction of the nuclear magnetic
moment is opposite to the nuclear spin and that a level with
higher total angular momentum ($F=1$) is below the level with the
lower value ($F=0$). That is different from the atomic hyperfine
structure of atoms where the nuclear magnetic moment is directed
along the nuclear spin (hydrogen, deuterium, muonium etc.).}
\begin{equation}
f_{\rm HFS}^{\rm D}(2s) = - 1083\,354\,981(9)\;{\rm Hz}
\end{equation}
and partly because various higher-order effects scales with the
nuclear charge $Z$ as $Z^5$ or $Z^6$ and thus are enhanced very
much for the helium ion ($Z=2$) in comparison with hydrogen and
deuterium ($Z=1$).

\subsection{The $1s$ hyperfine interval in muonium}

An alternative opportunity to verify precision theory for the
hyperfine effects is to study pure leptonic atoms. One of them is
muonium, a bound system of a positive muon and an electron. In
contrast to the hydrogen atom, the nucleus, a muon, is is not
influenced by the strong interaction. The strong interaction
enters only the system through hadronic vacuum polarization, which
sets an ultimate limit on any QED tests with muonium.
Uncertainties for QED itself and for the hadronic effects are
presented in \cite{strong,PRep1}. Muonium has metrological
interest due to determination of $\alpha$, $m_\mu/m_e$,
$\mu_\mu/\mu_p$ etc \cite{codata}.

\subsection{The $1s$ hyperfine interval in positronium}

Another atomic system of pure leptonic nature is positronium. The
nucleus, a positron, is a very light one. As a result, various
recoil effects, which are crucially important in advanced HFS
theory, are enhanced and therefore critical QED tests can be
performed with a relatively low experimental accuracy (see, e.g.,
\cite{PRep1} for detail).

\section{Summary of HFS tests of bound state QED}

Most recent results for precision test of bound state QED
involving the hyperfine structure of light hydrogen-like atoms are
summarized in Table~\ref{T:exp}. All tests presented are
competitive and the theoretical accuracy is limited by our ability
to calculate higher order radiative, recoil and radiative-recoil
effects (see review \cite{PRep1} for detail). Theory and
experiment are generally in good agreement, which is important for
metrological applications and confirms the reliability of
experimental and theoretical methods.

\begin{table}[hbtp]
\begin{center}
\begin{tabular}{lcccc}
\hline
 Atom  & ~~~~~~Exp.~~~~~~ & ~~~~~~Theor~~~~~~ &
~~~$\Delta/\sigma$~~~  \\
 & [kHz] & [kHz]  &    \\
 \hline
{H, $D_{21}$}  & 49.13(13) & 48.955(3)  & 1.4 \\
H,  $D_{21}$  &  48.53(23) &  & -1.8  \\
H,  $D_{21}$  &  49.13(40) & & 0.4  \\
{D, $D_{21}$}  & 11.280(56) & 11.312\,8(3) & -0.6  \\
D, $D_{21}$  &  11.16(16) &  & -1.0 \\
{$^3$He$^+$, $D_{21}$}~~~  &-1\,189.979(71) &-1\,190.14(6) &1.7 \\
$^3$He$^+$, $D_{21}$ & -1\,190.1(16) &  &  0.0  \\
\hline {Mu, $1s$} & 4\,463\,302.78(5) & 4\,463\,302.88(55)& -0.18 \\
Ps, $1s$ & 203\,389\,100(740)  & 203\,391\,700(500) & -2.9\\
Ps, $1s$ & 203\,397\,500(1600) & & -2.5 \\
\hline
\end{tabular}
\end{center}
\caption{Comparison of experiment and theory of hyperfine
structure in light hydrogen-like atoms. The reference can be found
in review \cite{PRep1}. \label{T:exp}}
\end{table}

\section*{Acknowledgement}

This work was supported in part by RFBR (grant \# 06-02-16156) and
DFG (grant GZ 436 RUS 113/769/0-2).


\begin{thebibliography}{99}

\bibitem{PRep1}
S.\ G.\ Karshenboim, ``Precision physics of simple atoms: QED
tests, nuclear structure and fundamental constants,'' \emph{Phys.
Rep.} {vol.\ 422}, pp.\ 1--63, 2005.

\bibitem{PRep2}
M.\ I.\ Eides,  H.\ Grotch and V.\ A.\ Shelyuto, ``Theory of light
hydrogenlike atoms,'' \emph{Phys. Rep.} vol.\ 342, pp.\ 63--261,
2001.

\bibitem{codata}
P.\ J.\ Mohr and B.\ N.\ Taylor, ``CODATA recomended values of the
fundamental physical constants: 2002 ,'' \emph{Rev. Mod. Phys.,}
vol. 77, pp. 1--107, 2005.

\bibitem{1s2s}
M. Niering, R. Holzwarth, J. Reichert, P. Pokasov, T. Udem, M.
Weitz, T. W. H\"ansch, P. Lemonde, G. Santarelli, M. Abgrall, P.
Laurent, C. Salomon, and A. Clairon, ``Measurment of the hydrogen
$1s-2s$ transition frequency by phase coherent comparison with a
microwave cesium fountain clock,'' \emph{Phys. Rev. Lett.,} vol.
84, pp. 5496--5499, 2000;\\ M. Fischer, N. Kolachevsky, M.
Zimmermann, R. Holzwarth, T. Udem, T. W. H\"ansch, M. Abgrall, J.
Grünert, I. Maksimovic, S. Bize, H. Marion, F. P. Dos Santos, P.
Lemonde, G. Santarelli, P. Laurent, A. Clairon, C. Salomon, M.
Haas, U. D. Jentschura, and C. H. Keitel, ``New limits on the
drift of fundamental constants from laboratory measurments,''
\emph{Phys. Rev. Lett.,} vol. 92, 230802, 2004.

\bibitem{H2s}
N.\ Kolachevsky, M.\ Fischer, S.\ G.\ Karshenboim and T.\ W.\
H\"{a}nsch,``High-Precision Optical Measurement of the $2s$
Hyperfine Interval in Atomic Hydrogen,'' \emph{Phys. Rev. Lett.,}
vol. 92, 033003, 2004.

\bibitem{D2s}
N.\ Kolachevsky, P.\ Fendel, S.\ G.\ Karshenboim and T.\ W.\
H\"{a}nsch, ``$2s$ hyperfine structure of atomic deuterium,''
\emph{Phys. Rev. A,} vol. 70, 062503, 2004.

\bibitem{JETP} S. G. Karshenboim, N. N. Kolachevsky, V. G. Ivanov, M. Fischer, P. Fendel, T. W.
H\"ansch, ``2s Hyperfine splitting in light hydrogen-like atoms:
Theory and experiment,'' \emph{JETP,} vol. 102, pp. 367 -- 379,
2006.

\bibitem{H2s_old} J.\
W.\ Heberle, H.\ A.\ Reich, and P.\ Kusch,``Hyperfine structure of
the metastable hydrogen atom,'' \emph{Phys. Rev.} vol. 101, pp.
612--620, 1956.

\bibitem{D2s_old}
H.\ A.~Reich, J.\ W.~Heberle, and P.~Kusch, ``Hyperfine structure
of the metastable deuterium atom,'' \emph{Phys. Rev.,} vol. 104,
pp. 1585--1592, 1956.

\bibitem{H2s_new}
N.\ E.\ Rothery and E.\ A.\ Hessels,``Measurment of the $2s$
atomic hydrogen hyperfine interval,'' \emph{Phys. Rev. A,} vol.
61, 044501, 2000.

\bibitem{d21first}
S.\ G.\ Karshenboim, ``2s Hyperfine Structure in Hydrogen Atom and
Helium-3 Ion,''. In ``Hydrogen atom: Precision physics of simple
atomic systems.,'' Ed. by S. G. Karshenboim et al., (Springer,
Berlin, Heidelberg, 2001) pp. 335--343.

\bibitem{th2s} S.\ G.\ Karshenboim and V.\ G.\ Ivanov,
``Hyperfine structure in hydrogen and helium ion,'' \emph{Phys.
Lett. B,} vol. 524, pp.\ 259--264, 2002;\\ S.\ G.\ Karshenboim and
V.\ G.\ Ivanov, ``Hyperfine structure of the ground and first
excited states in light hydrogen-like atoms and high-precision
tests of QED,'' \emph{Euro. Phys. J. D,} vol. 19, pp. 13--23,
2002.

\bibitem{th2s1}
S.\ G.\ Karshenboim and V.\ G.\ Ivanov, ``Improved theoretical
prediction for the 2s hyperfine interval in helium ion,''
\emph{Can. J. Phys.,} vol. 83, pp.\ 1063--1069, 2005.

\bibitem{newnumb} V. A. Yerokhin, A. N. Artemyev,
V. M. Shabaev, and G. Plunien, ``All-order results for the
one-loop QED correction to the hyperfine structure in light H-like
atoms,'' \emph{Phys. Rev. A,} vol. 72, 052510, 2005.

\bibitem{1sL1} M. Weitz, A. Huber,
F. Schmidt-Kaler, D. Leibfried, W. Vassen, C. Zimmermann, K.
Pachucki, T. W. H\"ansch, L. Julien, and F. Biraben, ``Precision
measurement of the 1S ground-state Lamb shift in atomic hydrogen
and deuterium by frequency comparison,'' \emph{Phys. Rev. A,} vol.
52, pp. 2664--2681, 1995

\bibitem{1sL2} D.
J. Berkeland, E. A. Hinds, and M. G. Boshier, ``Precise Optical
Measurement of Lamb Shifts in Atomic Hydrogen,'' \emph{Phys. Rev.
Lett.,} vol. 75, pp. 2470--2473, 1995.

\bibitem{1sL3} S. Bourzeix, B. de Beauvoir, F. Nez, M. D.
Plimmer, F. de Tomasi, L. Julien, F. Biraben, and D. N. Stacey,
``High Resolution Spectroscopy of the Hydrogen Atom: Determination
of the 1S Lamb Shift,'' \emph{Phys. Rev. Lett.,} vol. 76, pp.
384--387, 1996.

\bibitem{He2s}
M.\ H.\ Prior and E.\ C.\ Wang, ``Hyperfine structure of the $2s$
state of ${}^3He^+$,'' \emph{Phys. Rev. A,} vol.\ 16, pp.\ 6--18,
1977.

%%%%%%%%%%%%%%%

\bibitem{strong}
A.\ Czarnecki, S.\ I.\ Eidelman, and S.\ G.\ Karshenboim,
``Muonium hyperfine structure and hadronic effects,''  \emph{Phys.
Rev. D,} vol. {65}, 053004, 2002;\\ S.\ G.\ Karshenboim and V.\
A.\ Shelyuto, ``Hadronic vacuum polarization contribution to the
muonium hyperfine splitting,'' \emph{Phys. Lett. B,} vol. {517},
pp.\ 32--36, 2001;\\
 S.\ I.\ Eidelman, S.\ G.\ Karshenboim, and V.\ A.\
Shelyuto, ``Hadronic effects in leptonic systems: muonium
hyperfine structure and anomalous magnetic moment of muon,''
\emph{Can. J. Phys.,} vol.\ 80, pp. 1297--1303, 2002.

\end{thebibliography}
\end{document}